\documentclass[conference]{IEEEtran}
\IEEEoverridecommandlockouts

\usepackage{amsmath,amssymb,amsfonts}

\usepackage{algorithm}
\usepackage[noend]{algpseudocode} 

\usepackage{graphicx}
\usepackage{subcaption}
\usepackage{booktabs}
\usepackage{tabularx}
\usepackage{makecell}
\usepackage{braket}
\usepackage{cite}
\usepackage{xcolor}

\usepackage[hidelinks]{hyperref}

\algrenewcommand\algorithmicrequire{\textbf{Input:}}
\algrenewcommand\algorithmicensure{\textbf{Output:}}
\hypersetup{draft}

\usepackage[top=0.74in, bottom=1in, left=0.75in, right=0.75in]{geometry}

\hypersetup{
    colorlinks=true, 
    linkcolor=black,  
    citecolor=black, 
    filecolor=black, 
    urlcolor=black    
}

\def\BibTeX{{\rm B\kern-.05em{\sc i\kern-.025em b}\kern-.08em
    T\kern-.1667em\lower.7ex\hbox{E}\kern-.125emX}}
\begin{document}

\title{Recursive QAOA for Interference-Aware Resource Allocation in Wireless Networks}

\author{
\IEEEauthorblockN{
    Kuan-Cheng Chen\IEEEauthorrefmark{2},
    Hiromichi Matsuyama\IEEEauthorrefmark{3},
    Wei-hao Huang \IEEEauthorrefmark{3},
    Yu Yamashiro \IEEEauthorrefmark{3}
}
\IEEEauthorblockA{\IEEEauthorrefmark{2}\textit{J-ij Europe Ltd}, Rutherford Appleton Labortory, Harwell Campus, OX11 0QX, UK}
\IEEEauthorblockA{\IEEEauthorrefmark{3}\textit{Jij Inc.}, 3-3-6 Shibaura, Minato-ku, Tokyo, 108-0023, Japan}
}


\maketitle

\begin{abstract}
Discrete radio resource management problems in dense wireless networks are naturally cast as quadratic unconstrained binary optimization (QUBO) programs but are difficult to solve at scale. We investigate a quantum–classical approach based on the Recursive Quantum Approximate Optimization Algorithm (RQAOA), which interleaves shallow QAOA layers with variable elimination guided by measured single- and two-qubit correlators. For interference-aware channel assignment, we give a compact QUBO/Ising formulation in which pairwise interference induces same-channel couplings and one-hot constraints are enforced via quadratic penalties (or, optionally, constraint-preserving mixers). Within RQAOA, fixing high-confidence variables or relations reduces the problem dimension, stabilizes training, and concentrates measurement effort on a shrinking instance that is solved exactly once below a cutoff. On simulated instances of modest size, including a four-user, four-channel example, the method consistently returns feasible assignments and, for the demonstrated case, attains the global optimum. These results indicate that recursion can mitigate parameter growth and feasibility issues that affect plain QAOA, and suggest a viable pathway for near-term quantum heuristics in wireless resource allocation.

\end{abstract}


\section{Introduction}
\label{sec:intro}

Radio resource management (RRM) under stringent interference, capacity, and latency constraints remains a central challenge for 5G and emerging 6G systems \cite{bennis2018ultrareliable}. Core RRM tasks—such as channel assignment, user association, and discrete power control—are typically formulated as large-scale combinatorial optimization problems whose decision variants are NP-hard \cite{xu2021survey}. Classical solution strategies include exact mathematical programming methods (e.g., branch-and-bound and branch-and-cut for integer linear programs), as well as heuristic and metaheuristic algorithms such as graph coloring–based approaches and simulated annealing. Although these methods perform well across many operational regimes, their scalability can degrade significantly as network density increases, interference becomes highly nonlocal, or constraints introduce tight couplings across time, frequency, and space \cite{kaur2021machine}. These limitations motivate the exploration of alternative computational paradigms that natively handle discrete structure, support parallelizable primitives, and integrate effectively with probabilistic inference—capabilities increasingly exemplified by emerging quantum optimization algorithms \cite{abbas2024challenges,chen2025quantum-ci}.

The Quantum Approximate Optimization Algorithm (QAOA) is a prominent variational approach designed for binary optimization via an Ising or QUBO representation \cite{farhi2014quantum}. QAOA alternates between cost and mixer unitaries, with tunable angles optimized by a classical or quantum-enhanced optimizer\cite{chen2025learning}, depending on the implementation. QAOA offers a scalable framework for solving NP-hard combinatorial problems arising in applications such as wireless channel assignment\cite{chen2025resource,kim2021heuristic,zhao2024quantum}, traffic routing\cite{azad2022solving,fitzek2024applying,huang2025solving}, supply chain logistics\cite{azzaoui2021quantum}, scheduling\cite{ajagekar2022hybrid,kurowski2023application}, and portfolio optimization\cite{rebentrost2024quantum,buonaiuto2023best}, making it particularly well-suited for near-term quantum devices\cite{zhou2020quantum,chen2024noise,sakai2025transferring}. Despite promising empirical results on diverse combinatorial problems, plain QAOA faces two practical limitations for wireless resource allocation: (i) penalty-based constraint handling can require careful coefficient tuning and may yield infeasible samples, and (ii) the number of variational parameters and effective circuit depth rise with problem size, complicating training and execution\cite{blekos2024review,abbas2024challenges}. Recursive QAOA (RQAOA)~\cite{patel2024reinforcement} is a hybrid strategy that mitigates scaling challenges in QAOA by iteratively reducing the problem size. After each QAOA run, strongly correlated spin pairs are identified from measured expectation values, fixed accordingly, and eliminated to form a smaller effective instance. This variable-elimination loop improves feasibility on near-term hardware and reduces circuit resources while largely preserving solution quality~\cite{gulbahar2025majority}. However, RQAOA does not explicitly address coefficient heterogeneity, as problem weights remain encoded in the Ising Hamiltonian throughout the recursion.

We tailor RQAOA to \textit{interference‑aware resource allocation} in wireless networks, focusing on channel assignment with optional capacity and one‑hot constraints. The problem is expressed as an QUBO model where (i) interference costs are pairwise couplings between user–channel variables and (ii) assignment and capacity constraints are imposed via quadratic penalties or structure‑preserving mixers. Within the RQAOA loop, term‑wise expectation estimates guide variable fixing, consistency maintenance, and final exact classical search on the reduced instance.

This paper makes the following contributions:

\begin{itemize}
    \item \textbf{Problem-to-Ising mapping for wireless RRM.} We present a clean, extensible QUBO encoding for interference-aware channel allocation that supports one-hot user constraints and per-channel capacities, and discuss coefficient scaling to balance feasibility and optimality.

    \item \textbf{RQAOA specialization for network constraints.} We adapt recursive variable-fixing rules to the wireless setting, including correlation-based pair selection, recovery of eliminated indices, and feasibility repair consistent with assignment structure.

    \item \textbf{Software workflow and open implementation.} We use \texttt{Qamomile}\cite{huang2025qamomile}, a quantum optimization software, to provide a reproducible pipeline from wireless instances to quantum circuits and expectation estimation, suitable for both simulators and near-term devices.

    \item \textbf{Empirical evaluation against classical baselines.} On synthetic and benchmark topologies with geometric (pathloss-based) and graph-induced interference, we study solution quality, feasibility rate, and runtime as a function of users, channels, and constraint tightness, comparing RQAOA with  Integer Linear Programming (exact for small instances), greedy/coloring heuristics.
\end{itemize}

Taken together, our results and analysis position RQAOA as a practical quantum--classical heuristic for wireless resource allocation: recursion reduces instance size and sharpens structure, while QAOA layers capture nonlocal interference patterns that are difficult for purely greedy schemes. Beyond channel assignment, the same methodology extends to joint user--AP association, discrete power/beam selection, and time--frequency scheduling by expanding the decision variable set and reusing the recursive elimination machinery.

The remainder of the paper is organized as follows. Section~\ref{sec:preliminaries} formulates interference-aware resource allocation and derives the QUBO representation. Section~\ref{sec:preliminaries} reviews QAOA and details the recursive scheme, including expectation estimation and variable-fixing policies. Section~\ref{sec:experiments} presents experimental results and comparisons. Section~\ref{sec:conclusion} discusses limitations and opportunities---hardware noise, coefficient tuning, and conclusion.

\section{Preliminaries}\label{sec:preliminaries}

This section reviews the Quadratic Unconstrained Binary Optimization (QUBO) optimization formalism\cite{glover2018tutorial}, the QAOA, and its RQAOA. We also establish notation for interference-aware resource allocation in wireless networks and show how the problem is encoded as an Ising Hamiltonian solved by RQAOA.

\subsection{Ising and QUBO Formulations}
We consider binary optimization in two equivalent forms: the QUBO in variables $x_i\in\{0,1\}$,
\begin{equation}
\min_{x\in\{0,1\}^n}\; f(x)=x^\top Q x + q^\top x + c,
\label{eq:qubo}
\end{equation}
and the Ising model in spin variables $s_i\in\{-1,+1\}$,
\begin{equation}
\min_{z\in\{-1,+1\}^n}\; E(s)=\sum_{i<j} J_{ij}\, z_i z_j + \sum_i h_i\, z_i + \mathrm{const}.
\label{eq:ising}
\end{equation}
The two are related by $x_i=(1-z_i)/2$, which induces an affine mapping between $(Q,q,c)$ and $(J,h,\mathrm{const})$. In the quantum setting, the Ising objective becomes a diagonal Hamiltonian
\begin{equation}
H_C \;=\; \sum_{i<j} J_{ij}\, Z_i Z_j + \sum_i h_i\, Z_i \;+\; \mathrm{const}\, \mathbb{I},
\label{eq:cost-ham}
\end{equation}
where $Z_i$ is the Pauli-$Z$ operator acting on qubit $i$.

\subsection{QAOA Basics}

We consider binary optimization problems encoded in Ising form
\begin{equation}
E(\boldsymbol{z}) \;=\; \mathrm{const} + \sum_i h_i z_i + \sum_{i<j} J_{ij} z_i z_j,
\qquad z_i \in \{-1,+1\},
\label{eq:ising}
\end{equation}
which induce a diagonal cost Hamiltonian \(H_C\) in the computational basis. More generally, QAOA optimizes a classical objective function \(C:\{0,1\}^n \to \mathbb{R}\) by lifting it to
\begin{equation}
H_C \;=\; \sum_{x\in\{0,1\}^n} C(x)\,\ket{x}\!\bra{x},
\label{eq:qaoa-cost}
\end{equation}
and pairing it with a mixing Hamiltonian \(H_M\).

The associated unitaries are
\begin{equation}
U_C(\gamma) = e^{-i\gamma H_C}, 
\qquad
U_M(\beta) = e^{-i\beta H_M},
\label{eq:qaoa-unitaries}
\end{equation}
and the depth-\(p\) QAOA ansatz prepares the variational state
\begin{equation}
\ket{\psi_p(\boldsymbol{\gamma},\boldsymbol{\beta})}
= \prod_{\ell=1}^{p} U_M(\beta_\ell)\, U_C(\gamma_\ell)\,\ket{\psi_0},
\label{eq:qaoa-state}
\end{equation}
where \(\ket{\psi_0}\) is an efficiently preparable initial state (e.g., \(\ket{+}^{\otimes n}\) or a constraint-satisfying product state).

Unless otherwise stated, we adopt the canonical transverse-field mixer
\begin{equation}
H_M \;=\; \sum_{i=1}^n X_i,
\qquad
U_M(\beta)=\prod_{i=1}^n e^{-i\beta X_i},
\label{eq:mixer}
\end{equation}
which is appropriate for unconstrained binary optimization problems such as MaxCut. More general mixer constructions—including \(RY\)-type and constraint-preserving mixers (e.g., ring-\(XY\) mixers)—can be employed to restrict evolution to feasible subspaces in constrained settings~\cite{wang2020xy}.

The variational parameters \((\boldsymbol{\gamma},\boldsymbol{\beta})=(\gamma_1,\dots,\gamma_p;\,\beta_1,\dots,\beta_p)\) are optimized to minimize
\begin{equation}
F_p(\boldsymbol{\gamma},\boldsymbol{\beta})
=\bra{\psi_p(\boldsymbol{\gamma},\boldsymbol{\beta})} H_C \ket{\psi_p(\boldsymbol{\gamma},\boldsymbol{\beta})},
\label{eq:objective}
\end{equation}
using a classical outer loop. When \(H_C\) is an Ising-form Hamiltonian, \(U_C(\gamma)\) decomposes into parallel single-qubit \(Z\) rotations and two-qubit \(ZZ\) interactions, while \(U_M(\beta)\) consists of parallel single-qubit rotations; the circuit depth per layer is therefore determined by the interaction graph of \(H_C\).

For sufficiently large depth \(p\), QAOA approaches digitized adiabatic evolution under \(H(s)=(1-s)H_M+sH_C\), whereas at finite depth it remains a shallow, hardware-compatible variational ansatz. Estimating \(F_p\) requires repeated measurement of commuting Pauli terms in \(H_C\); noise effects can be mitigated using standard techniques such as readout calibration, symmetry verification, and zero-noise extrapolation.

\subsection{Recursive QAOA}
\label{RQAOA}
\begin{algorithm}[!b]
\caption{Recursive QAOA}
\label{alg:RQAOA}
\begin{algorithmic}[1]
\Require Ising instance $(h, J, \mathrm{const})$, depth $p$, mixer $U_M$, initializer $\ket{\psi_0}$, cutoff $n_\mathrm{cutoff}$, threshold $\tau$, shots $M$
\Ensure Spin assignment $\hat{\mathbf{z}} \in \{-1,+1\}^n$
\State $S \gets \{1,\dots,n\}$ \Comment{active indices}
\State $\mathcal{R} \gets \emptyset$ \Comment{elimination record}
\While{$|S| > n_\mathrm{cutoff}$}
  \State Optimize QAOA on $H_C$ to obtain $(\boldsymbol{\beta}, \boldsymbol{\gamma})$
  \State Estimate $\langle Z_i\rangle$ and $\langle Z_i Z_j\rangle$ for $i,j\in S$
  \State Compute scores $s_i=|\langle Z_i\rangle|$, $s_{ij}=|\langle Z_i Z_j\rangle|$
  \State Select $t^\star$ with maximum score above $\tau$
  \If{$t^\star = (i,j)$} \Comment{pair decimation}
     \State $\sigma \gets \mathrm{sign}(\langle Z_i Z_j\rangle)$
     \State Record $z_j=\sigma z_i$, eliminate $j$, update $(h, J, \mathrm{const})$
  \ElsIf{$t^\star = i$} \Comment{single-spin fixing}
     \State $\sigma \gets \mathrm{sign}(\langle Z_i\rangle)$
     \State Record $z_i=\sigma$, eliminate $i$, update $(h, J, \mathrm{const})$
  \Else
     \State \textbf{break}
  \EndIf
\EndWhile
\State Solve reduced instance exactly
\State Back-substitute eliminated variables via $\mathcal{R}$
\State \Return $\hat{\mathbf{z}}$
\end{algorithmic}
\end{algorithm}

\emph{Recursive QAOA} augments QAOA with a \emph{variable-elimination loop}. After each QAOA optimization, we estimate selected observables—typically single- and pairwise $Z$-moments,
\(\langle Z_i\rangle\) and \(\langle Z_iZ_j\rangle\)—from which we infer high-confidence relations (signs and correlations) that allow us to \emph{fix} variables and shrink the instance. The recursion reduces circuit width and depth requirements while often improving constraint satisfaction and optimizer stability.

Let \(E(\boldsymbol{z})\) be as above. When we decide to fix 
\(z_k=\sigma\in\{-1,+1\}\) (e.g., using \(\sigma=\mathrm{sign}\,\langle Z_k\rangle\)), 
the reduced Ising instance on the remaining spins is  
\begin{align}
\mathrm{const}' &= \mathrm{const} + h_k \sigma, \\
h_i' &= h_i + J_{ik}\sigma, \quad (i \neq k), \\
J_{ij}' &= J_{ij}, \quad (i,j \neq k).
\end{align}
When we merge a strongly (anti)correlated pair via 
\(z_j=\sigma\,z_i\) with \(\sigma=\mathrm{sign}\,\langle Z_iZ_j\rangle\), 
we eliminate \(z_j\) and update  
\begin{align}
\mathrm{const}' &= \mathrm{const} + J_{ij}\sigma, \\
h_i' &= h_i + \sigma h_j, \\
J_{ik}' &= J_{ik} + \sigma J_{jk}, \quad (k \notin \{i,j\}),
\end{align}
while removing index \(j\).  
These transformations follow from substituting 
\(z_k\) or \(z_j\) and using \(z_i^2=1\).

We form a score for each candidate term (e.g., \(|\langle Z_i\rangle|\) and \(|\langle Z_iZ_j\rangle|\)), apply a threshold \(\tau\) (or a statistical bound via Hoeffding/Chernoff using $M$ shots), and preferentially eliminate the highest-scoring pair terms (which remove one variable and rewire couplings). The recursion stops once the active set size is at most \(n_{\min}\), where a small exact solver (brute force, ILP) is feasible. Eliminated relations are stored and later back-substituted to recover a full assignment.

(i) The problem dimension shrinks monotonically, so successive QAOA calls act on fewer qubits and admit shallower effective circuits. 
(ii) Pairwise decimation redistributes weight into effective couplings that more explicitly encode the instance structure. 
(iii) When constraint-preserving mixers are employed (e.g., ring-\(XY\) mixers for one-hot user–channel blocks), and/or when moderate penalty coefficients are used, feasibility is maintained by construction: the quantum evolution remains within the constraint manifold while recursive elimination removes weakly informative or ambiguous variables. 
Although RQAOA does not guarantee monotonic energy improvement at each elimination step, it empirically yields robust, high-quality solutions on interference-limited allocation tasks.


\section{Wireless Model: Interference-Aware Resource Allocation}
We focus on interference-aware channel assignment in a wireless network with \(U\) users and \(C\) orthogonal channels. Let \(x_{u,c}\in\{0,1\}\) indicate whether user \(u\) is assigned to channel \(c\). Co-channel transmissions give rise to mutual interference, which we quantify through an \emph{interference weight} \(w^{(c)}_{u,v}\ge 0\) capturing the cost incurred when users \(u\) and \(v\) share the same channel \(c\) (e.g., as a function of pathloss, spatial proximity, or empirical interference statistics). The objective is to allocate channels so as to minimize the aggregate co-channel interference subject to per-user assignment constraints:

\begin{align}
\min_{x}\quad 
& \sum_{c=1}^{C} \sum_{1\le u<v\le U} w^{(c)}_{u,v}\, x_{u,c} x_{v,c} \label{eq:wireless-obj}\\
\text{s.t.}\quad 
& \sum_{c=1}^{C} x_{u,c} = 1,\quad \forall u \in [U], \label{eq:onehot}\\
& \sum_{u=1}^{U} x_{u,c} \le K_c,\quad \forall c \in [C], \label{eq:cap}
\end{align}
where \([U]\triangleq\{1,2,\ldots,U\}\) and \([C]\triangleq\{1,2,\ldots,C\}\) denote the user and channel index sets, respectively. 
Constraint~\eqref{eq:onehot} enforces exactly one channel per user (one-hot), and~\eqref{eq:cap} optionally enforces the per-channel capacity \(K_c\).

We adopt a standard penalty method to produce a QUBO. Let $A>0$ penalize \eqref{eq:onehot} and $B\ge 0$ penalize \eqref{eq:cap}. The penalized objective is
\begin{align}
\min_x\quad 
& \sum_{c} \sum_{u<v} w^{(c)}_{u,v} x_{u,c} x_{v,c}
+ A \sum_{u}\Big(\sum_{c} x_{u,c}-1\Big)^2 \nonumber\\
&\quad + B \sum_{c}\Big(\sum_{u} x_{u,c} + \sum_{\ell=0}^{L_c-1} 2^\ell y_{c,\ell}-K_c\Big)^2 .
\label{eq:qubo-wireless}
\end{align}
Index the assignment variables by \(i(u,c)=(u-1)C+c\), yielding \(n_x=UC\) binary variables \(x_{u,c}\). If per-channel capacities \eqref{eq:cap} are enforced via a quadratic penalty, we introduce binary slack variables \(y_{c,\ell}\) such that \(s_c=\sum_{\ell=0}^{L_c-1}2^\ell y_{c,\ell}\) with \(L_c=\lceil\log_2(K_c+1)\rceil\), and penalize \(\sum_u x_{u,c}+s_c=K_c\); the total number of binary variables becomes \(n=n_x+\sum_{c=1}^C L_c\). Expanding the squared penalties in~\eqref{eq:qubo-wireless} then yields a QUBO of the form~\eqref{eq:qubo} with explicit linear and quadratic coefficients. Applying the standard spin mapping \(x_i=(1-z_i)/2\) (and analogously for \(y_{c,\ell}\)) produces the Ising parameters \((J,h,\mathrm{const})\) in~\eqref{eq:ising} and the diagonal cost Hamiltonian~\eqref{eq:cost-ham} used by QAOA.

\subsection{Mixers}
This work primarily uses the transverse-field mixer \eqref{eq:mixer}, which is hardware-friendly and general. Constraint-preserving mixers (e.g., one-hot or ring-exchange mixers per user to conserve $\sum_c x_{u,c}$) can reduce penalty burden and improve landscapes; we discuss them as design variants, but they are not required for the RQAOA pipeline.

\begin{figure*}[!t]
    \centering
    \includegraphics[width=\linewidth]{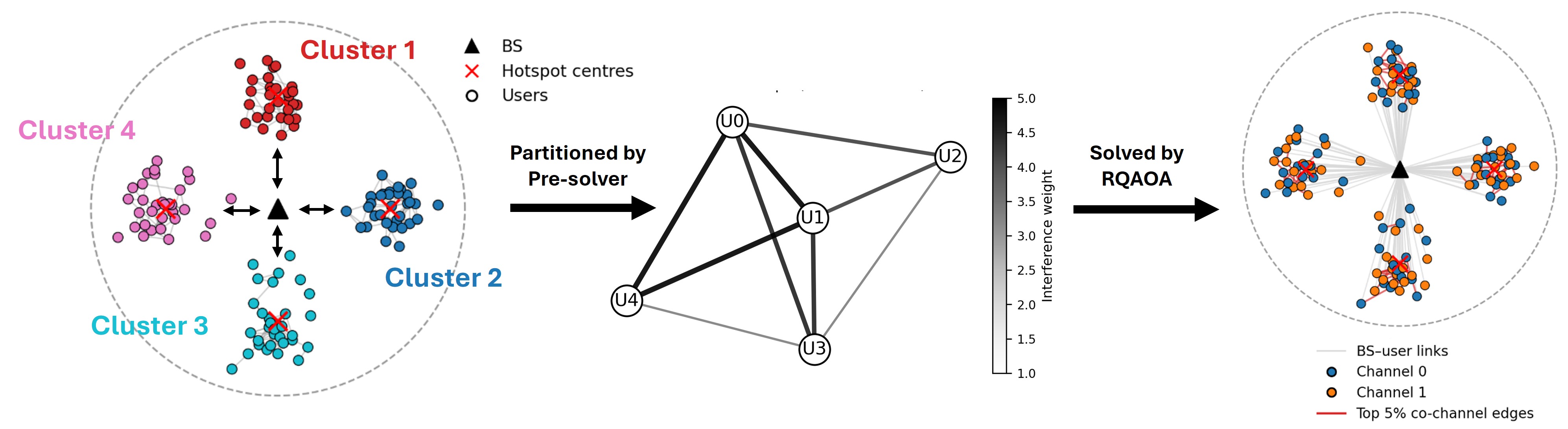}
    \caption{Classical pre-solving and graph reductions for RQAOA. 
    Spatial user clusters (left) are aggregated into a reduced interference 
    graph (center), which is then processed by RQAOA to generate channel 
    assignments (right).}
    \label{fig:rqaoa-demo}
\end{figure*}

\subsection{Applying RQAOA to the Wireless Ising}

Given the wireless Ising instance derived from~\eqref{eq:qubo-wireless}, we apply RQAOA following the generic variable-elimination procedure described in Sec.~\ref{RQAOA}, with diagnostics and fixing rules tailored to the problem structure. In each recursion, we estimate selected single- and pairwise observables, focusing on correlations that reflect (i) one-hot constraints for each user across channels and (ii) interference between users sharing the same channel. High-confidence correlations are used to fix variables or merge spins, yielding a reduced Ising instance with updated couplings.

Constraint handling is integrated by design: one-hot feasibility is enforced either via constraint-preserving mixers (e.g., ring-\(XY\) blocks over user–channel variables) or via moderate penalty coefficients in the Ising formulation, while recursive elimination removes weakly informative or ambiguous assignments. The recursion terminates once the remaining instance is sufficiently small for exact solution, after which eliminated variables are recovered by back-substitution. This problem-aware instantiation of RQAOA progressively sharpens interference structure and maintains feasibility throughout the reduction.

\subsection{Output Decoding and Feasibility}
The final spin configuration $s$ is mapped to $x$ via $x_i=(1-s_i)/2$, reshaped to $X\in\{0,1\}^{U\times C}$ with $X_{u,c}=x_{i(u,c)}$. Feasibility is checked by $\sum_c X_{u,c}=1$ and $\sum_u X_{u,c}\le K_c$. If soft-penalty violations occur, a lightweight repair step (e.g., reassigning ties to the least-loaded channels) can be applied, though in our experience appropriate $(A,B)$ and recursion typically yield feasible $X$ directly.

\subsection{Classical Pre-solving and Graph Reductions}
\label{sec:presolve}

To further improve scalability, we insert a lightweight \emph{classical pre-solver} in front of the RQAOA loop (shown in Fig.~\ref{fig:rqaoa-demo}). The pre-solver applies deterministic QUBO reductions and, optionally, heuristic variable fixing to shrink the Ising instance before any quantum calls. Let
\[
E(\boldsymbol{z}) \;=\; \mathrm{const} + \sum_i h_i z_i + \sum_{i<j} J_{ij} z_i z_j,\qquad z_i\in\{-1,+1\},
\]
denote the Ising form derived from the wireless QUBO.

\paragraph{Deterministic reductions.}
We first apply structural simplifications that preserve global optimality:
\begin{itemize}
    \item \emph{Isolated spins.} If a spin $z_i$ has no couplings ($J_{ij}=0$ for all $j$), then its optimal value is determined by the sign of $h_i$; we fix $z_i=\mathrm{sign}(-h_i)$, update the constant term, and remove $i$.
    \item \emph{Dominance (persistency) tests.} If a spin satisfies
    \begin{equation}
      |h_i| \;\ge\; \sum_{j\neq i} |J_{ij}|,
      \label{eq:persistency}
    \end{equation}
    then in any global minimizer we must have $z_i^\star = \mathrm{sign}(-h_i)$. We fix $z_i$ accordingly and update $(\mathrm{const}, h, J)$ as in the RQAOA elimination step.
    \item \emph{Wireless-specific pruning.} In the channel-assignment model, users or channels that have become effectively single-choice after capacity and interference pruning can be fixed deterministically. For example, if for user $u$ only one channel $c^\star$ remains feasible, we set $x_{u,c^\star}=1$ and $x_{u,c}=0$ for $c\neq c^\star$, then eliminate the corresponding spins.
\end{itemize}
These rules are iterated until no further change occurs, yielding a reduced Ising instance $(h',J',\mathrm{const}')$ over an active index set $S_0\subseteq\{1,\dots,n\}$ and an elimination record $\mathcal{R}_0$.

\paragraph{Heuristic freezing from classical local search}
Optionally, we exploit fast classical heuristics to identify \emph{frozen} spins. We run a local-search solver (e.g., greedy descent or simulated annealing) on the original QUBO and collect a set of low-energy solutions $\{\boldsymbol{z}^{(k)}\}_{k=1}^K$. For each spin we compute an empirical magnetization
\begin{equation}
m_i \;=\; \frac{1}{K}\sum_{k=1}^K z_i^{(k)}.
\end{equation}
If $|m_i|\ge \tau_{\mathrm{freeze}}$ for a threshold $\tau_{\mathrm{freeze}}\in(0,1)$ (e.g., $0.9$), we treat $z_i$ as effectively frozen and fix $z_i=\mathrm{sign}(m_i)$, again updating $(h,J,\mathrm{const})$ and recording the relation in $\mathcal{R}_0$.

\paragraph{Integration with RQAOA}
The overall hybrid pipeline becomes
\begin{enumerate}
    \item Construct the wireless QUBO and corresponding Ising parameters \((h,J,\mathrm{const})\).
    \item Apply the classical pre-solver to obtain \((h',J',\mathrm{const}')\) on a reduced index set \(S_0\), and record eliminations \(\mathcal{R}_0\).
    \item Run RQAOA (Alg.~\ref{alg:RQAOA}) on the reduced instance until the cutoff \(n_{\min}\) is reached, yielding an additional elimination record \(\mathcal{R}_\mathrm{RQAOA}\) and a small core problem.
    \item Solve the core exactly and back-substitute all recorded relations \((\mathcal{R}_0 \cup \mathcal{R}_\mathrm{RQAOA})\) to recover a full assignment in the original user--channel variables.
\end{enumerate}
In practice, this pre-solving stage reduces the number of spins passed to RQAOA, shortens quantum circuit width, and concentrates RQAOA effort on the structurally hardest part of the wireless instance, while preserving the global structure of the solution.
Crucially, in the large-scale setting we apply RQAOA only to a fixed-size core (selected by an interference score), so the number of qubits and QAOA calls is bounded by \((U_{\mathrm{core}}, n_{\mathrm{cutoff}})\) and does not scale with the total user count \(U\).

\section{Experiments}
\label{sec:experiments}

\begin{figure}[!b]
  \centering
  \begin{subfigure}{0.8\columnwidth}
    \centering
    \includegraphics[width=\linewidth]{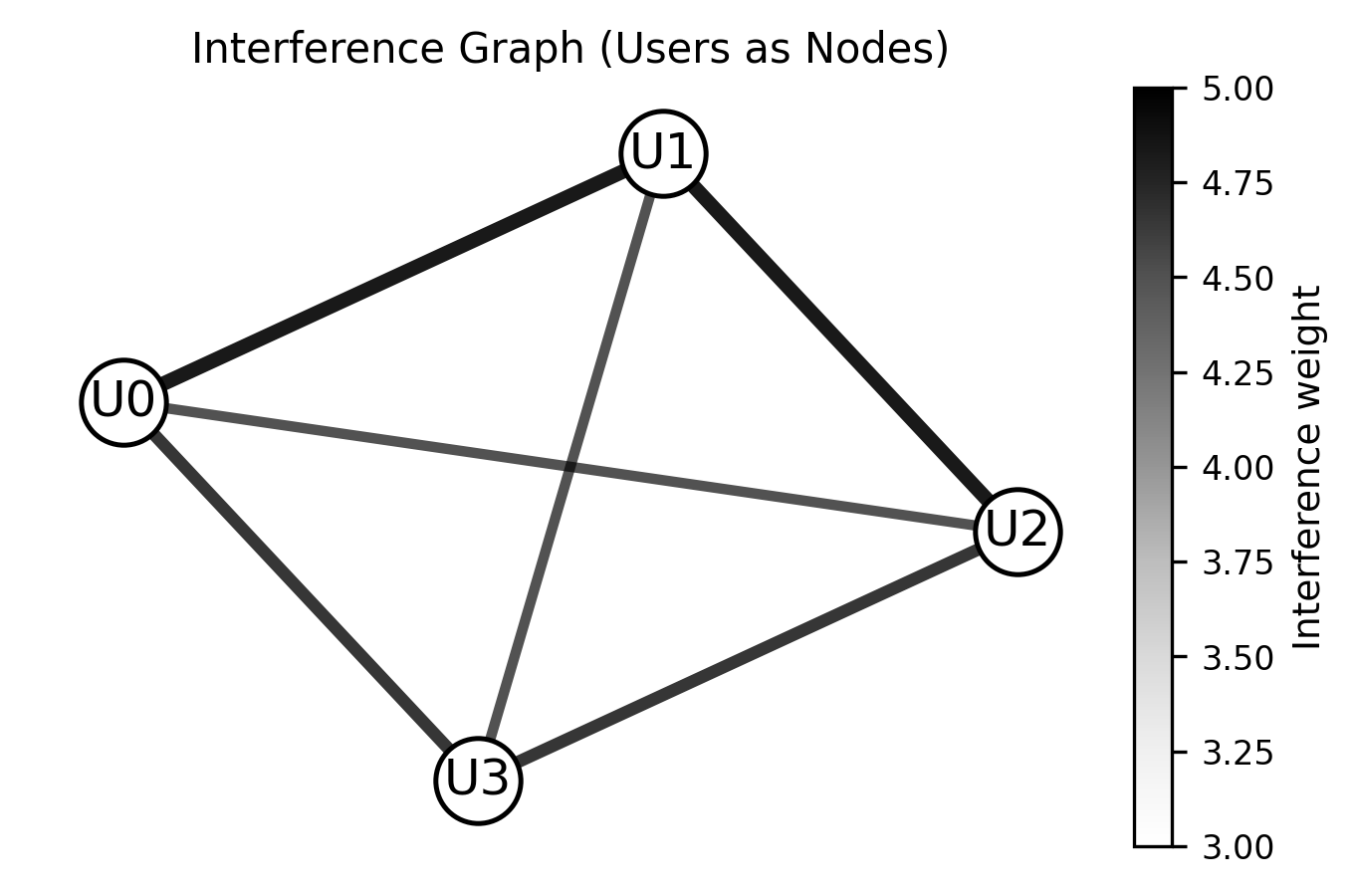}
    \caption{Interference graph}
  \end{subfigure}\\[6pt]
  \begin{subfigure}{0.8\columnwidth}
    \centering
    \includegraphics[width=\linewidth]{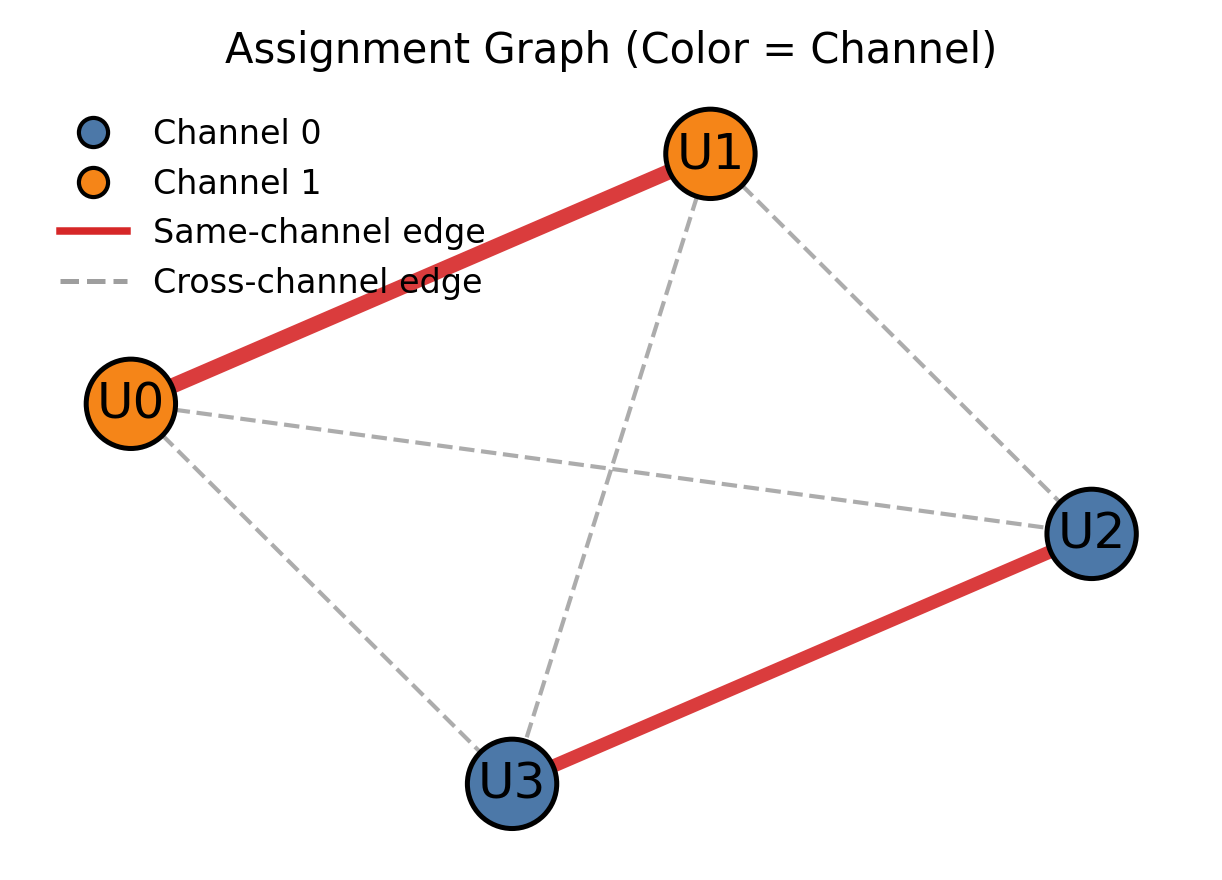}
    \caption{RQAOA channel assignment}
  \end{subfigure}
  \caption{Interference topology and RQAOA-derived channel assignment for a \(U=4, C=4\) instance.}
  \label{fig:wireless-figs}
\end{figure}

We first benchmark RQAOA on a wireless channel-assignment instance cast as a QUBO model. There are \(U\) users and \(C\) orthogonal channels; each user must be assigned exactly one channel while minimizing pairwise interference. Let \(x_{u,c}\!\in\!\{0,1\}\) indicate that user \(u\) uses channel \(c\), with one-hot constraint \(\sum_c x_{u,c}=1\). Interference between users \(i\) and \(j\) is specified by a symmetric nonnegative weight \(w_{ij}\). With penalty strength \(A>0\), the diagonal cost Hamiltonian is
\begin{equation}
\begin{aligned}
H_C &= A \sum_{u=0}^{U-1}\!\Big(\sum_{c=0}^{C-1} x_{u,c}-1\Big)^2
\;+\; \sum_{i<j}\sum_{c=0}^{C-1} w_{ij}\, x_{i,c} x_{j,c},\\[-2pt]
x_{u,c} &= \tfrac{1-Z_{u,c}}{2},
\end{aligned}
\label{eq:wireless-H}
\end{equation}
where \(Z_{u,c}\) acts on the qubit representing the pair \((u,c)\). Equation~\eqref{eq:wireless-H} compactly encodes the linear, quadratic, and constant contributions. After the spin mapping, the Hamiltonian takes the standard Ising form
\(
H_C=\mathrm{const}
      +\sum h_{u,c} Z_{u,c}
      +\sum J_{(u,c),(v,d)} Z_{u,c}Z_{v,d},
\)
with coefficients determined by \(A\) and the interference weights \(w_{ij}\).

At each recursion step, depth-\(p\) QAOA is applied to the current reduced Hamiltonian to estimate the single- and two-body correlators \(\langle Z_i\rangle\) and \(\langle Z_i Z_j\rangle\). The term with largest magnitude is then fixed: for a single index \(i\), the spin is set to \(s_i=\mathrm{sign}\big(\langle Z_i\rangle\big)\); for a pair \((i,j)\), the relation \(s_i s_j=\mathrm{sign}\big(\langle Z_i Z_j\rangle\big)\) is imposed. Substituting these relations reduces the problem size and produces a constant energy shift. The process repeats until the number of remaining spins is at most \(n_{\!\mathrm{cutoff}}\), at which point the reduced instance is solved exactly by full enumeration and the solution is lifted by propagating the fixed relations back through the recursion.

As an illustrative example, we consider \(U=4\) users and \(C=4\) channels with integer interference weights \(w_{ij}\!\in\!\{0,\dots,5\}\) (fixed random seed). The penalty parameter is set to \(A=10\) so that any violation of the one-hot constraints is energetically suppressed, and QAOA is run with depth \(p=1\) and cutoff \(n_{\!\mathrm{cutoff}}=6\). Performance is quantified by (i) feasibility rate, i.e., the fraction of samples obeying \(\sum_c x_{u,c}=1\) for all \(u\); (ii) achieved objective \(C(z)\); and (iii) approximation ratio relative to the global optimum obtained by exhaustive enumeration. For this demonstration instance, RQAOA returns a feasible assignment and matches the globally optimal objective.

Figure~\ref{fig:wireless-figs} summarizes the demonstration instance and the recovered solution. 
Fig.~\ref{fig:wireless-figs}(a) shows the interference graph for \(U=4\) users, where edge thickness and grayscale encode the interference weight \(w_{ij}\). 
For each channel \(c\), this induces an Ising coupling \(J_{(i,c),(j,c)} = w_{ij}\) in the cost Hamiltonian, while one-hot constraints are enforced via the penalty term \(\sum_c x_{u,c} = 1\). 
Fig.~\ref{fig:wireless-figs}(b) displays the RQAOA-derived assignment (\(p=1\), \(n_{\!\mathrm{cutoff}}=6\), \(A=10\)), with node colors indicating the selected channel per user. 
Solid edges denote user pairs assigned to the same channel and thus contributing an interference cost proportional to \(w_{ij}\), whereas dashed edges correspond to cross-channel pairs that incur no such penalty. 
For this seeded instance, the resulting configuration is feasible and attains the global optimum.

\begin{table}[!t]
\centering
\caption{Summary of mixer types considered in RQAOA experiments.}
\label{tab:mixers}
\begin{tabular}{p{2.1cm}p{5.8cm}}
\toprule
\textbf{Mixer} & \textbf{Description} \\
\midrule
Ring-XY & XY interactions in a ring topology within each user group; preserves one-hot constraints with minimal depth~\cite{he2023alignment}. \\
Clique-XY & XY interactions on all pairs in each group (complete graph); fastest state mixing at the cost of higher depth~\cite{rieffel2020xy}. \\
Matching-XY & XY interactions applied in disjoint pairings over two parallel rounds; one-hot preserving and depth efficient~\cite{rieffel2020xy}. \\
Star-XY & Single ``anchor'' qubit coupled to all others in its group; preserves one-hot constraints with moderate depth~\cite{kordonowy2025lie}. \\
RY & Single-qubit \(R_y\) rotations on all qubits; unconstrained mixing that may break one-hot structure unless penalties are strong~\cite{fuchs2024lx}. \\
RX & Single-qubit \(R_x\) rotations on all qubits; analogous to RY with a different rotation axis~\cite{fuchs2024lx}. \\
\bottomrule
\end{tabular}
\end{table}

For small problem sizes, several mixer families were investigated, as summarized in Table~\ref{tab:mixers}. Constraint-preserving XY mixers (ring, clique, matching, star) maintain the one-hot structure exactly and therefore guarantee feasibility at the wavefunction level, whereas single-qubit RX/RY mixers explore a larger portion of the Hilbert space but rely on the penalty term \(A\) in~\eqref{eq:wireless-H} to discourage invalid assignments. Unless otherwise stated, the large-scale wireless experiments in the following subsection employ the standard RX mixer for simplicity, as feasibility is enforced by the QUBO penalties and the recursion itself is agnostic to the particular mixer family.

\begin{figure}[!t]
  \centering
  \includegraphics[width=\columnwidth]{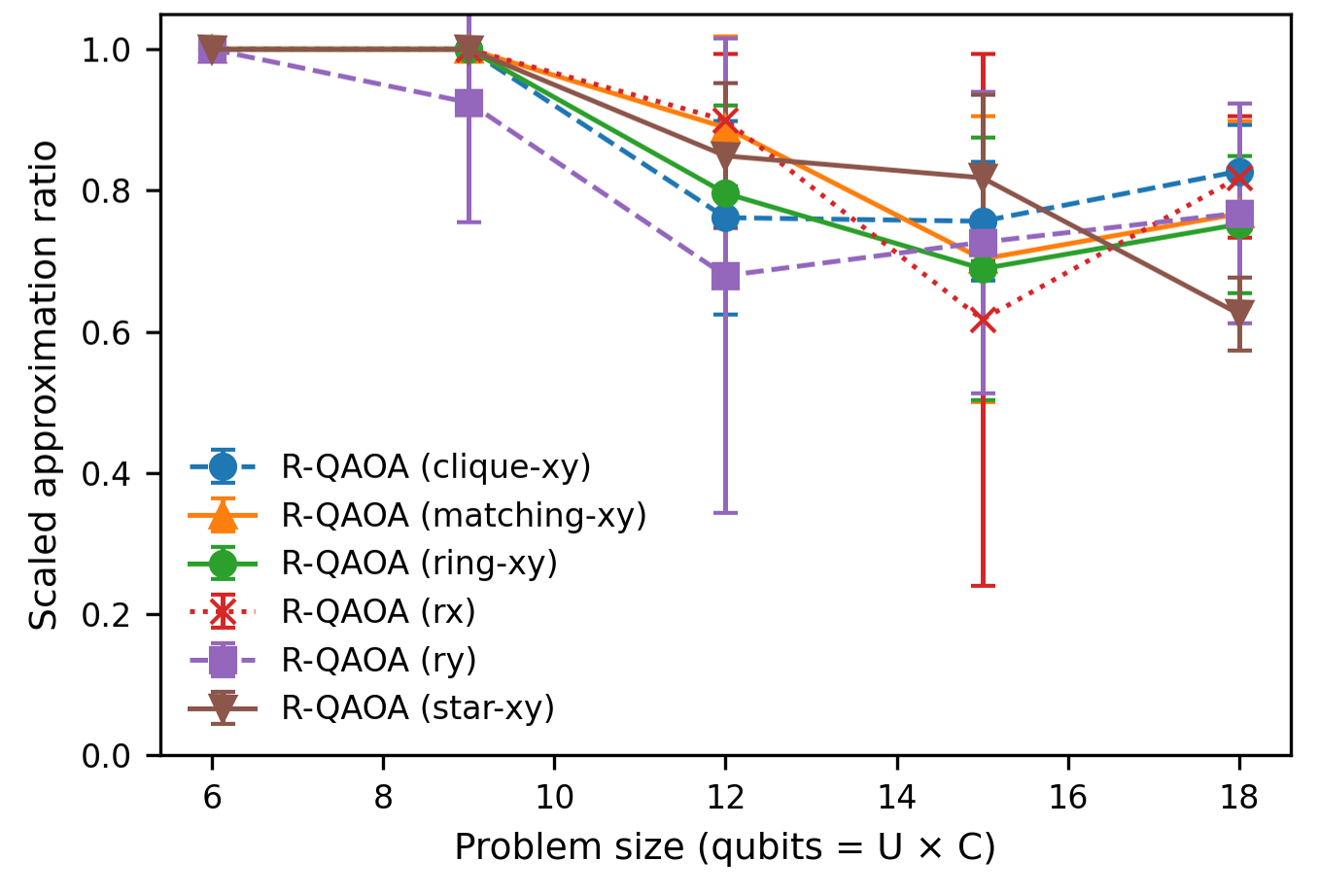}
  \caption{Scaled approximation ratio (mean $\pm$ std, five seeds) versus problem size for RQAOA with a ring-XY mixer (\(C=3\), \(p=2\), \(n_{\min}=8\)), normalized between each instance’s best and worst feasible energies. RQAOA remains near-optimal at small sizes and around \(0.7\text{--}0.8\) at larger sizes, with feasibility preserved.}
  \label{fig:RQAOA-size-sweep}
\end{figure}

Figure~\ref{fig:RQAOA-size-sweep} reports the mean and standard deviation of the scaled approximation ratio—the energy normalized between each instance’s best and worst feasible energies—over five random seeds per size. RQAOA is essentially optimal for \(n\in\{6,9\}\) qubits and remains competitive for \(n\in\{12,15,18\}\), where the average ratio stabilizes around \(0.7\text{--}0.8\) while the variance increases as the interference structure and parameter landscape become more complex. Feasibility remains at \(100\%\) across all runs due to the constraint-preserving ring-XY mixer, and recursion keeps the circuit width bounded as variables are eliminated. The mild performance degradation with increasing size is consistent with fixed shallow depth and nonconvex training; improvements are expected from higher depth \(p\), warm-started parameters across recursion steps, adaptive elimination criteria, and more advanced measurement strategies.

\subsection{Large-Scale Wireless Hotspot Benchmarks}
\label{subsec:large-scale-results}

We now move from qubit-scale toy problems to a more realistic wireless hotspot scenario, using the wireless channel-assignment model of~\eqref{eq:wireless-H} with stochastic user locations and log-normal shadowing. For each network size 
\[
U \in \{16, 32, 64, 128, 256, ... , 32768\},
\]
ten independent hotspot topologies are generated. In all cases there are \(C=2\) orthogonal channels, and a penalty parameter \(A\) is chosen automatically from the empirical distribution of interference weights to ensure that violating the one-hot constraints is energetically unfavourable. For each realised instance, three solvers are applied: a full-graph greedy heuristic, a QAOA-based pipeline with a fixed core of \(U_{\mathrm{core}}=10\) highly interfering users followed by greedy extension, and the proposed RQAOA pipeline operating on the same core with recursive variable elimination before extension. For larger instances, we first extract this quantum-sized core by ranking users via an interference score (e.g., weighted degree) and restricting the QUBO/Ising to the induced subgraph; the remaining users are then assigned sequentially by a low-complexity greedy rule that places each user on the channel minimizing its incremental interference with already-assigned users while respecting capacities. All quantum subroutines are run on a state-vector simulator with identical circuit depth and classical optimizer settings across problem sizes.

\begin{figure}[!t]
  \centering
  \includegraphics[width=\columnwidth]{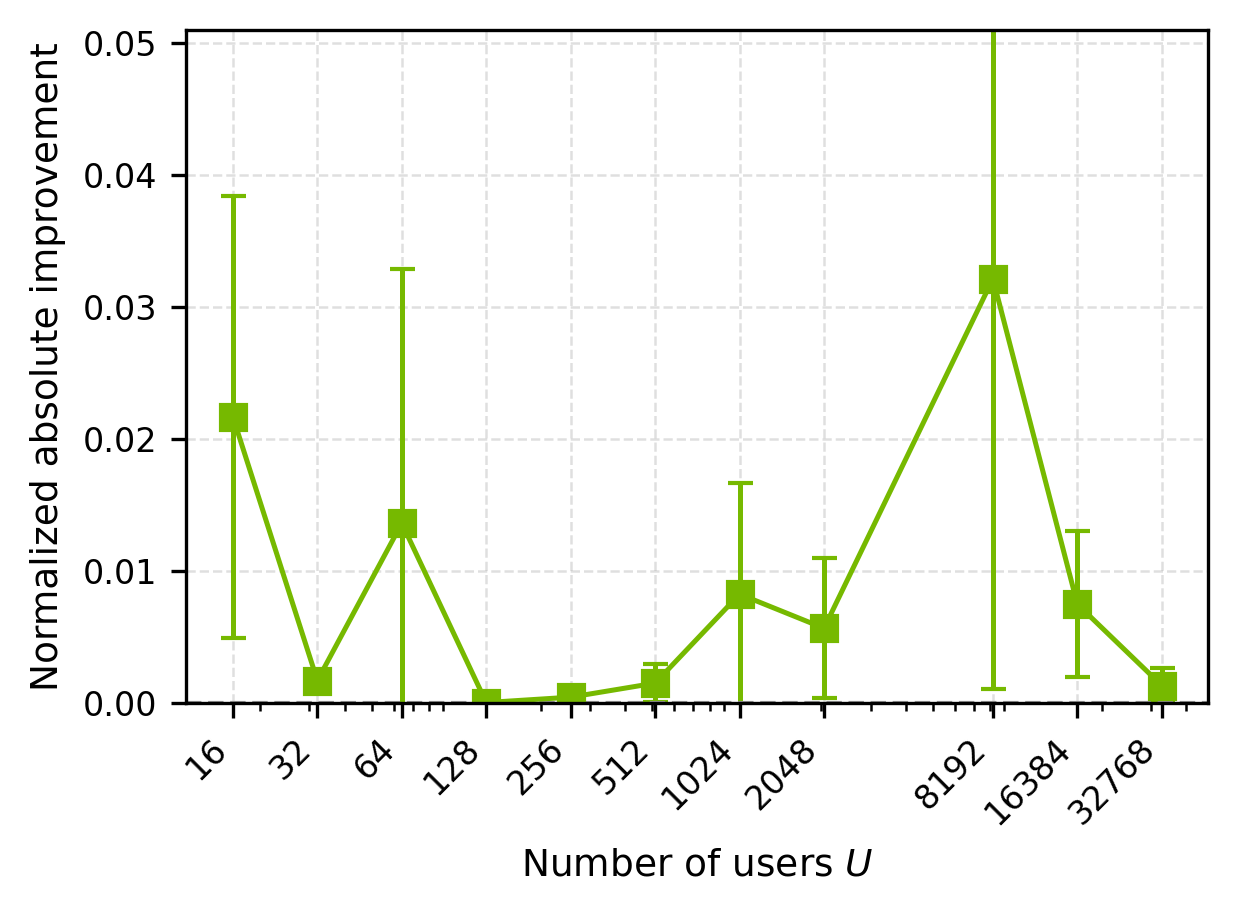}
  \caption{Normalized absolute deviation between R-QAOA and greedy baseline,
  as a function of the number of users $U$ (mean $\pm$ std over random hotspot topologies).}
  \label{fig:rqaoa_gap_abs}
\end{figure}

\begin{figure}[!t]
  \centering
  \includegraphics[width=\columnwidth]{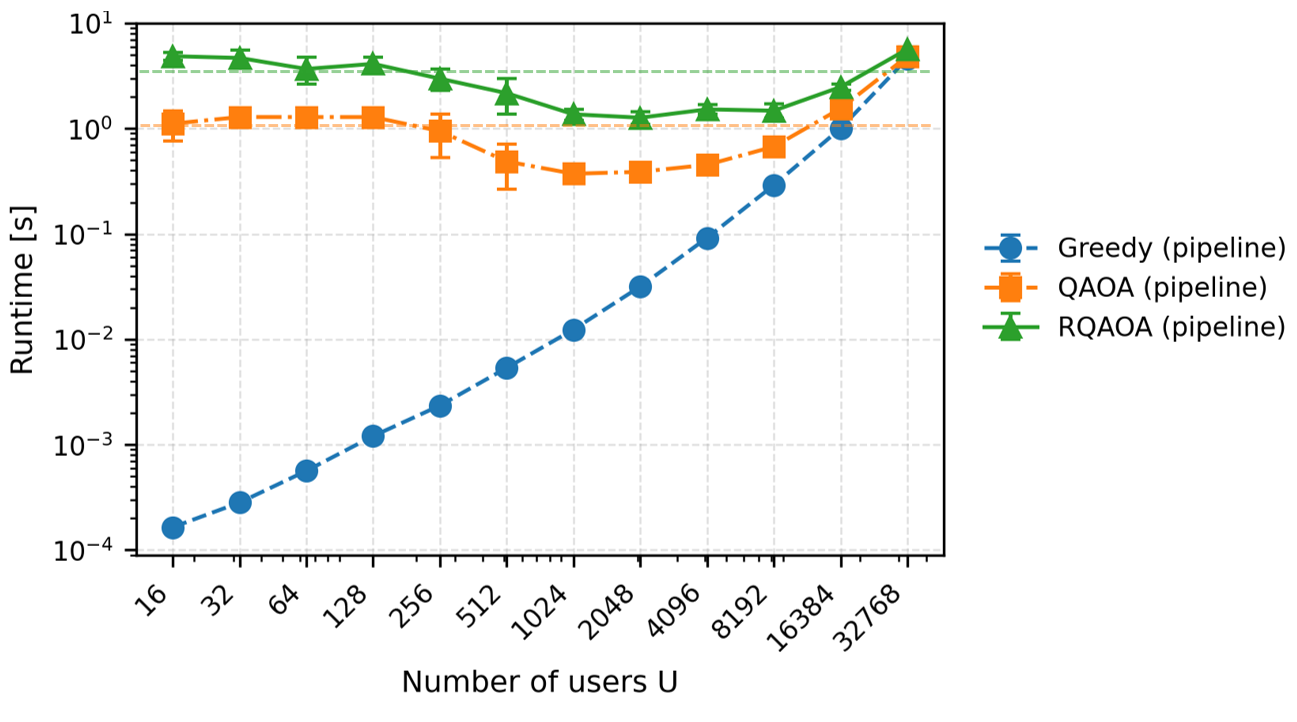}
  \caption{End-to-end pipeline runtime versus number of users $U$ (log scale) for the greedy heuristic, QAOA pipeline, and R-QAOA pipeline (mean $\pm$ std over random hotspot topologies).}
  \label{fig:pipeline_runtime}
\end{figure}

Figure~\ref{fig:rqaoa_gap_abs} shows the normalized deviation
\begin{equation}
  \Delta_{\mathrm{norm}}(U)
  = \frac{\bigl|C_{\mathrm{R{\text -}QAOA}}(U)-C_{\mathrm{Greedy}}(U)\bigr|}
         {C_{\mathrm{Greedy}}(U)},
  \label{eq:normalized_gap_abs}
\end{equation}
between the interference cost of the R-QAOA pipeline and that of the classical greedy baseline, averaged over the random topologies for each $U$. Across the entire range from \(U=16\) to \(U=32768\), the mean deviation remains small, typically at or below a few percent. For small and medium-scale networks (\(U \leq 512\)), the average $\Delta_{\mathrm{norm}}(U)$ is well below $10^{-2}$, demonstrating that the recursive quantum core combined with classical extension can reproduce the behaviour of the full-graph greedy heuristic with high fidelity in terms of aggregate interference. For larger networks, the deviation remains modest: around \(U=8192\) the mean deviation is on the order of \(3\times 10^{-2}\), with larger error bars (computed over five independent repetitions) reflecting increased instance-to-instance variability in the underlying interference graphs. The deviation then decreases again for \(U=16384\) and \(U=32768\). These results indicate that, for the considered hotspot models, R-QAOA closely tracks a strong classical baseline over more than three orders of magnitude in network size, even though the quantum subsystem is restricted to a fixed ten-user core.

The corresponding end-to-end runtime behaviour is reported in Fig.~\ref{fig:pipeline_runtime}. The greedy heuristic exhibits approximately linear growth with \(U\), consistent with operating on the full interference graph: runtimes increase from sub-millisecond at \(U=16\) to a few seconds at \(U=32768\). The QAOA and RQAOA pipelines follow a different trend because the quantum subproblem is a fixed-size core of \(U_{\mathrm{core}}=10\) users solved at fixed depth and shot budget, while the remaining users are assigned by a lightweight greedy extension. Consequently, for small and moderate \(U\) the runtime is dominated by the constant-cost quantum-core optimization and is nearly independent of the total network size. The gradual increase beyond \(U\approx 10^3\) is primarily attributable to classical overheads—core selection on larger graphs and extension over more users—with a secondary contribution from occasional increases in optimizer iterations when the selected core exhibits stronger and more heterogeneous couplings. Overall, bounding the circuit width by \(U_{\mathrm{core}}\) yields substantially milder scaling with \(U\) than a full-graph quantum optimizer.

Taken together, Figs.~\ref{fig:rqaoa_gap_abs} and~\ref{fig:pipeline_runtime} show that the proposed R-QAOA pipeline offers a favourable trade-off for interference-aware channel assignment in dense wireless deployments. The aggregate interference achieved by R-QAOA is essentially indistinguishable from that of a strong greedy heuristic, while the quantum subproblem remains of fixed size and the runtime overhead stays within one to two orders of magnitude of the purely classical baseline. From a telecommunication systems perspective, this suggests that embedding small, programmable quantum cores in the control plane via recursive hybrid schemes such as R-QAOA can provide quantum-enhanced scheduling capabilities without incurring prohibitive latency or scalability penalties.

\section{Conclusion}
\label{sec:conclusion}
We have presented a QUBO formulation of interference-aware channel assignment and tailored RQAOA to this problem class. The recursive loop—identifying large-magnitude correlators, fixing spins or pairwise relations, and algebraically reducing the Hamiltonian—progressively shrinks circuit width while promoting satisfaction of the per-user one-hot constraints; in noiseless simulations, RQAOA consistently produced feasible solutions and, for the illustrated instance, recovered the global optimum obtained by exhaustive search. Remaining limitations include the need to carefully scale penalty coefficients to balance feasibility against objective distortion, the reliance on state-vector simulators rather than noisy devices, and sensitivity of convergence to parameter initialization and the observable-selection policy. Future work will focus on systematic benchmarks against strong classical heuristics on large synthetic and geometry-derived topologies, hardware demonstrations with measurement-frugal and error-mitigated implementations, improved constraint-preserving mixers, and extensions to joint association, beam/power selection, and distributed multi-QPU realizations of the RQAOA pipeline for scalable wireless resource management\cite{burt2024generalised,burt2025multilevel,burt2025entanglement,chen2026adaptive}.

\section*{Acknowledgment}
The authors acknowledge the use of Jij's Quantum SDK, \texttt{Qamomile}~\cite{huang2025qamomile} (https://github.com/Jij-Inc/Qamomile).

\bibliographystyle{ieeetr}
\bibliography{references}

\end{document}